# Photoionization Effects During Galaxy Formation

MATTHIAS STEINMETZ

*Max–Planck–Institut für Astrophysik*
*Postfach 1523*
*85740 Garching, Germany*

## INTRODUCTION

In addition to gravitation, radiative cooling and feedback processes due to supernovae, radiative heating by photoionization might play a major role in the hydrodynamics of forming galaxies. In the first place, it provides an important feedback process on scales smaller than 1 kpc[1], which can be even more efficient than supernovae. Secondly, an external UV background field can significantly affect the collapse and cooling history of the gas: the photon background raises the gas temperature to a few $10^4$ K increasing the Jeans mass to $\approx 10^8 \, M_\odot$ in baryons[2]. On the other hand, the photoionization depletes the fraction of neutral hydrogen and helium and, therefore, reduces the cooling due to collisional excitation, the dominant cooling process for a gas of primordial composition. Compared to the case of no background radiation, the cooling time of the gas can be enlarged by 2 to 3 orders of magnitude[3].

It is widely believed, that quasars are responsible for the UV background. Hence, the UV spectra are usually assumed to be quasar like[3,4]:

$$I(\nu) = I_{-21} \left(\frac{\nu}{\nu_H}\right)^{-\alpha} \quad [10^{-21}\,\text{erg}/\text{cm}^2/\text{Hz}/\text{sec}/\text{sterad}] \quad (1)$$

where $\nu_H$ corresponds to the hydrogen ionization energy of 13.6 eV. For photoionization to be an important process, intensities $I_{-21}$ of $1-10$ and rather hard spectra ($\alpha < 2$) are required. Both requirements are within the limits imposed on the UV background by the Gunn–Peterson effect and the proximity effect[3], but possibly in conflict with the observed X–ray background.

The photoionization rate is proportional to the density, whereas the recombination rate is proportional to $\varrho^2$. The UV background is therefore most important at low densities ($< 10^{-2}\,\text{cm}^{-3}$). Indeed, even collisionless systems which collapses at redshift $z > 3$, which are typical for haloes less massive than $10^{11}\,M_\odot$, are too dense for photoionization to become important. Therefore, the importance of the UV background does not only critically depend on its spectrum and intensity, but also whether it becomes sufficiently strong at epochs when haloes leave the linear regime. Insufficient numerical resolution can lead to a completely different and even qualitatively wrong result: Due to the missing resolution the formation period of the smallest structures might be delayed to epochs when the photon field is strong enough to prevent collapse, whereas in a higher resolved simulation these objects are already collapsed and inert against photoionization. Most work so far was restricted to one–dimensional hydrodynamical simulations[5].

Figure 1: Distribution of gas particles at $z = 2.9$ for four models with identical initial conditions but different UV background fields. Every frame is 200 kpc across (in physical coordinates). Upper left: no UV background; upper middle: strong UV background ($\alpha = 1$, $I_{-21} = 10$) from the beginning; upper right: Strong UV background, exponentially switched on between $z = 8 - 5$; lower left: $\alpha = 1$, $I_{-21} = 1$ from the beginnning; lower middle: $\alpha = 5$, $I_{-21} = 10$ from the beginnning; lower right: $\alpha = 5$, $I_{-21} = 1$ from the beginnning. Absorption effects are included for the simulations shown in the lower three panels.

## Numerical Simulations

We have performed three-dimensional, high–resolution hydrodynamical simulations with *Smoothed Particle Hydrodynamics*. The fully time dependent ionization network equations are solved simultaneously on the hydrodynamical or cooling time step whichever is smaller. Effects due to collisional excitation and ionization, recombination as well compton heating and cooling with the microwave and UV background are taken into account. At the relevant densities below $10^{-3} - 10^{-2} \, \text{cm}^{-3}$, the cooling, ionization and recombination time scales are comparable or even longer than the dynamical time scale and a non equilibrium description is required.

The simulations start with a spherically symmetric volume at a redshift of $z = 25$ and contain a mass of $8 \, 10^{11} \, M_\odot$ assuming vacuum boundary conditions[6,7]. Small scale power is added according to a $b = 2$ biased CDM spectrum. The influence of modes with larger wavelengths is approximated by raising the mass of the sphere by 27% corresponding to a $2.5\,\sigma$ peak of a biased CDM spectrum on a mass scale of $8\ 10^{11}\,M_\odot$. The tidal field is approximated by setting the sphere initially into rigid rotation with a spin parameter of $\lambda = 0.05$. Since the sphere reaches turn around at $z \approx 5$ and is fully collapsed only at $z = 2$, the growth of structures is fairly well described up to redshifts $z \approx 2.5$. The simulations were done with 18000 particles in the dark matter and also in the gas ($\Omega_{\rm bary} = 0.1$), corresponding to individual masses of $4\ 10^6\,M_\odot$ and $3.6\ 10^7\,M_\odot$, respectively. The gravitational softening is 1 kpc for the dark matter and 0.5 kpc for the gas.

For the UV background, we assume a power law as given by Eq. (**1**) with $I_{-21} = 1$ or 10 and $\alpha = 1$ or 5. For some models, absorption effects were taken into account. In principle this requires a solution of the radiation transport equations and is still beyond the capabilities of the current generation of supercomputers. We, therefore, calculate the intensities modified by the spatially averaged opacities of the gas in the considered volume. The resulting spectrum is similar to those calculated by Madau[3] for photon energies between 13.6 and 100 eV, although it is noticeably stronger than those obtained by Cen and Ostriker[2]. Furthermore, in some models the UV background was not present from the beginning but was exponentially switched on or off over a redshift interval of about 3. A detailed description of the way the network equations are solved, and how absorption effects are taken into account are postponed to a forthcoming publication.

## RESULTS

Only in the case of an intense ($I_{-21} \gtrsim 1$) and hard ($\alpha \lesssim 2$) quasar like UV background, is the formation of small structures ($M \lesssim 3 \cdot 10^8\,M_\odot$) significantly suppressed (see also Fig. 1,2). Also the mass of the most massive halo ($2\ 10^{10}\,M_\odot$) can be reduced by 20%. However, such models are not compatible with the observed X–ray background, unless absorption effects are sufficiently strong[2] (which would, of course, again lower the photoionisation heating). In addition to small structures being depleted, cooling due to collisional excitation is completely suppressed. In contrast to simulations without a UV background, where the gas can cool efficiently and is kept at temperatures around $10^4$ K, in the case of a strong UV background energy is gained by gravitational collapse and cannot be radiated away. The gas temperature can reach the virial temperature of the halo, i.e. a few $10^6$ K for a Milky Way sized halo. As one might expect from the different density scaling of photoionisation and recombination described above, the effects of photoionisation are more prominent for a lower baryon fraction ($\Omega_{\rm bary} < 0.1$).

The influence of the UV background field is almost negligible in this simulations, if the spectrum becomes softer ($\alpha \gtrsim 2$) or less intense ($I_{-21} \lesssim 1$). The effect also becomes negligible, if the UV background is turned on too late when most of the structures with masses of $M \lesssim 3\ 10^8\,M_\odot$ have already collapsed (Fig. 1,2 lower left panel). For a biased CDM spectrum ($b = 2$) this corresponds to redshifts of $z \lesssim 7$. Switching off the radiation field at some point in the evolution, the gas cannot be held outside the dark haloes. It cools and collapses as soon as the background becomes weaker than $I_{-21} \approx 1$.

If absorption is taken into account, the quasar-like external background field is reduced up to 2 orders of magnitude in the range between 13.6 eV and 1 keV. Only in the most extreme model ($I_{21} = 10$, $\alpha = 1$) small scale objects are suppressed, but only if the intensity of the background radiation is already high at $z > 10$.

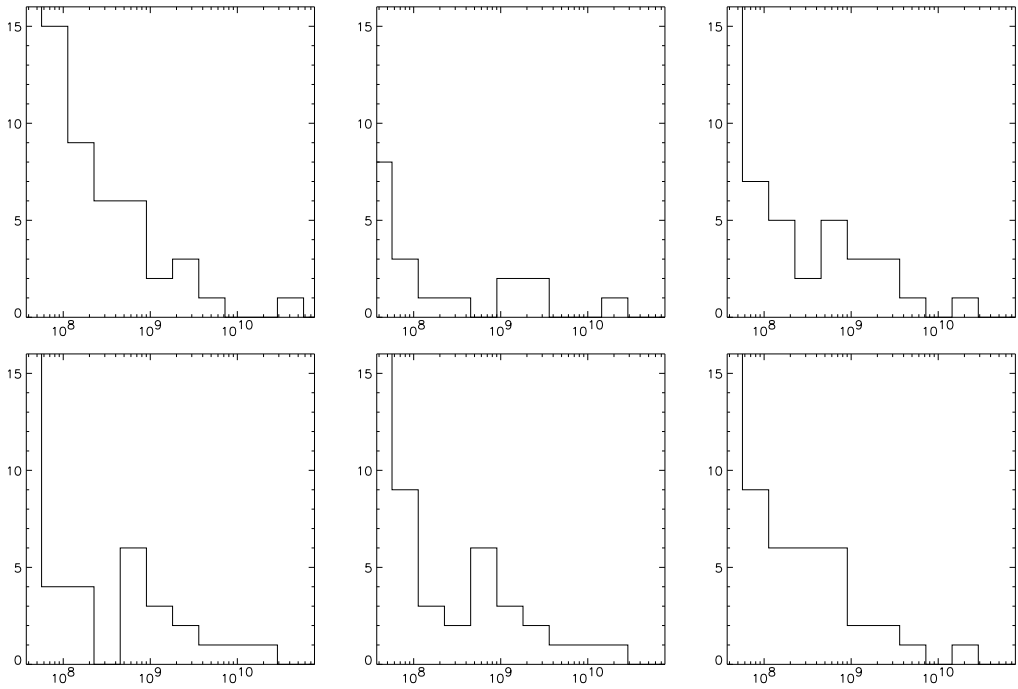

Figure 2: Number of groups $N$ of mass $M$ at $z = 2.9$ for the for models in Fig. 1.

In summary, extreme, if not unrealistic assumptions for the UV background must be made in order to significantly change the formation of galaxies more massive than a few $10^8\,M_\odot$.

## ACKNOWLEDGMENTS

It is a pleasure to thank Martin Haehnelt, Ewald Müller, Jerry Ostriker and Simon White fore many helpful discussions.